\pdfoutput=1

\documentclass[twocolumn,aps,prl,showpacs,superscriptaddress]{revtex4}

\usepackage{times}
\usepackage{graphicx}
\usepackage{epsfig}
\usepackage{amsfonts}
\usepackage{amsmath}
\usepackage{amssymb}
\usepackage{color}
\usepackage{multirow}
\usepackage[colorlinks=true,citecolor=blue,urlcolor=blue]{hyperref}

\begin{document}

%%%%%%%%%%%%%%%%%%%%%%%%%%%%%%%%%%%%%%%%%%%%%%%%%%%%%%%%%%%%%%%%%%%

\title{Nonlocality from Local Contextuality}

%bhliu@ustc.edu.cn,huxm@mail.ustc.edu.cn,chenjs10@mail.ustc.edu.cn,hyf@ustc.edu.cn,smhan@ustc.edu.cn,cfli@ustc.edu.cn,gcguo@ustc.edu.cn,adan@us.es

%%%%%%%%%%%%%%%%%%%%%%%%%%%%%%%%%%%%%%%%%%%%%%%%%%%%%%%%%%%%%%%%%%%

%Bi-Heng Liu bhliu@ustc.edu.cn
%Xiao-Min Hu huxm@mail.ustc.edu.cn
%Jiang-Shan Chen chenjs10@mail.ustc.edu.cn
%Yun-Feng Huang hyf@ustc.edu.cn
%Yong-Jian Han smhan@ustc.edu.cn
%Chuan-Feng Li cfli@ustc.edu.cn
%Guang-Can Guo gcguo@ustc.edu.cn
%Ad\'an Cabello adan@us.es

%%%%%%%%%%%%%%%%%%%%%%%%%%%%%%%%%%%%%%%%%%%%%%%%%%%%%%%%%%%%%%%%%%%

\author{Bi-Heng Liu}
\affiliation{Key Laboratory of Quantum Information, University of Science and Technology of China, Chinese Academy of Sciences, Hefei, 230026, People's Republic of China}
\affiliation{Synergetic Innovation Center of Quantum Information and Quantum Physics, University of Science and Technology of China, Hefei, 230026, People's Republic of China}

\author{Xiao-Min Hu}
\affiliation{Key Laboratory of Quantum Information, University of Science and Technology of China, Chinese Academy of Sciences, Hefei, 230026, People's Republic of China}
\affiliation{Synergetic Innovation Center of Quantum Information and Quantum Physics, University of Science and Technology of China, Hefei, 230026, People's Republic of China}

\author{Jiang-Shan Chen}
\affiliation{Key Laboratory of Quantum Information, University of Science and Technology of China, Chinese Academy of Sciences, Hefei, 230026, People's Republic of China}
\affiliation{Synergetic Innovation Center of Quantum Information and Quantum Physics, University of Science and Technology of China, Hefei, 230026, People's Republic of China}

\author{Yun-Feng Huang}
\affiliation{Key Laboratory of Quantum Information, University of Science and Technology of China, Chinese Academy of Sciences, Hefei, 230026, People's Republic of China}
\affiliation{Synergetic Innovation Center of Quantum Information and Quantum Physics, University of Science and Technology of China, Hefei, 230026, People's Republic of China}

\author{Yong-Jian Han}
\email{smhan@ustc.edu.cn}
\affiliation{Key Laboratory of Quantum Information, University of Science and Technology of China, Chinese Academy of Sciences, Hefei, 230026, People's Republic of China}
\affiliation{Synergetic Innovation Center of Quantum Information and Quantum Physics, University of Science and Technology of China, Hefei, 230026, People's Republic of China}

\author{Chuan-Feng Li}
\email{cfli@ustc.edu.cn}
\affiliation{Key Laboratory of Quantum Information, University of Science and Technology of China, Chinese Academy of Sciences, Hefei, 230026, People's Republic of China}
\affiliation{Synergetic Innovation Center of Quantum Information and Quantum Physics, University of Science and Technology of China, Hefei, 230026, People's Republic of China}

\author{Guang-Can Guo}
\affiliation{Key Laboratory of Quantum Information, University of Science and Technology of China, Chinese Academy of Sciences, Hefei, 230026, People's Republic of China}
\affiliation{Synergetic Innovation Center of Quantum Information and Quantum Physics, University of Science and Technology of China, Hefei, 230026, People's Republic of China}

\author{Ad\'{a}n Cabello}
\email{adan@us.es}
\affiliation{Departmento de F\'{\i}sica Aplicada II,Universidad de Sevilla,E-41012 Sevilla, Spain}

%%%%%%%%%%%%%%%%%%%%%%%%%%%%%%%%%%%%%%%%%%%%%%%%%%%%%%%%%%%%%%%%%%%

\date{\today}

%%%%%%%%%%%%%%%%%%%%%%%%%%%%%%%%%%%%%%%%%%%%%%%%%%%%%%%%%%%%%%%%%

\begin{abstract}
We experimentally show that nonlocality can be produced from single-particle contextuality by using two-particle correlations which do not violate any Bell inequality by themselves. This demonstrates that nonlocality can come from an {\em a priori} different simpler phenomenon, and connects contextuality and nonlocality, the two critical resources for, respectively, quantum computation and secure communication. From the perspective of quantum information, our experiment constitutes a proof of principle that quantum systems can be used simultaneously for both quantum computation and secure communication.
\end{abstract}

%%%%%%%%%%%%%%%%%%%%%%%%%%%%%%%%%%%%%%%%%%%%%%%%%%%%%%%%%%%%%%%%%

\pacs{03.65.Ta,03.65.Ud, 42.50.Xa}
%03.65.Ta: Foundations of quantum mechanics; measurement theory
%03.65.Ud: Entanglement and quantum nonlocality
%(e.g. EPR paradox, Bell's inequalities, GHZ states, etc.)
%42.50.Xa: Optical tests of quantum theory

\maketitle

%%%%%%%%%%%%%%%%%%%%%%%%%%%%%%%%%%%%%%%%%%%%%%%%%%%%%%%%%%%%%%%%%%%

{\em Introduction.---}Two famous ``no-go'' theorems prove that the predictions of quantum theory cannot be explained with hidden variables: Bell's theorem \cite{Bel64} states that they cannot be reproduced with local hidden variables (LHV) and the Bell-Kochen-Specker (BKS) theorem \cite{Spe60,Bel66,KS67} states that they cannot be explained by noncontextual hidden variables (NCHV). Recently, it has been recognized that each of these theorems is behind one of the resources that empower quantum information processing: Bell nonlocality is essential for device-independent secure communication \cite{Eke91,BHK05,ABG07} and BKS contextuality supplies the power for fault-tolerant universal quantum computation \cite{AB09,Rau13,HWV14,DGB15,RBD15}. This observation puts the problem of what is the relation between contextuality and nonlocality under a new perspective. In particular, it raises the question of whether single-particle contextuality and two-party nonlocality can coexist, so the same quantum system can provide both resources simultaneously. Surprisingly, the answer to this question is negative if we restrict ourselves to simple forms of nonlocality and single-particle contextuality as, in these cases, there are monogamies between them \cite{KCK14,JWG16,SR16} recently observed in experiments \cite{ZZL16}.

However, Kochen \cite{Koc70}, Stairs \cite{Sta78,Sta83}, and Heywood and Readhead \cite{HR83} noticed that the answer is different when single-particle contextuality is state independent. Then, contextuality can be converted into two-particle nonlocality by using Einstein-Podolsky-Rosen (EPR) correlations \cite{EPR35}. It follows that the conflict between quantum theory and LHV theories can be traced back to a conflict between quantum theory and NCHV theories for a single particle. In plain words, nonlocality can be produced from
single-particle contextuality by using two-particle correlations which do not violate any Bell inequality by themselves. This connects, in an
operational way, a fundamental physical phenomenon, nonlocality, with an {\em a priori} different phenomenon, single-particle contextuality,
providing a new perspective on the origin of nonlocality. Furthermore, from the perspective of quantum information, we see that the two
critical resources needed for, respectively, quantum computation and secure communication can both be simultaneously produced by the same system.

%%%%%%%%%%%%%%%%%%%%%%%%%%%%%%%%%%%%%%%%%%%%%%%%%%%%%%%%%%%%%%%%%
% Fig. 1
%%%%%%%%%%%%%%%%%%%%%%%%%%%%%%%%%%%%%%%%%%%%%%%%%%%%%%%%%%%%%%%%%

\begin{figure}[t]
\centering
\hspace{0.6cm}
\includegraphics[trim = 1.2cm 14cm 0 8.2cm,clip,width=8.6cm]{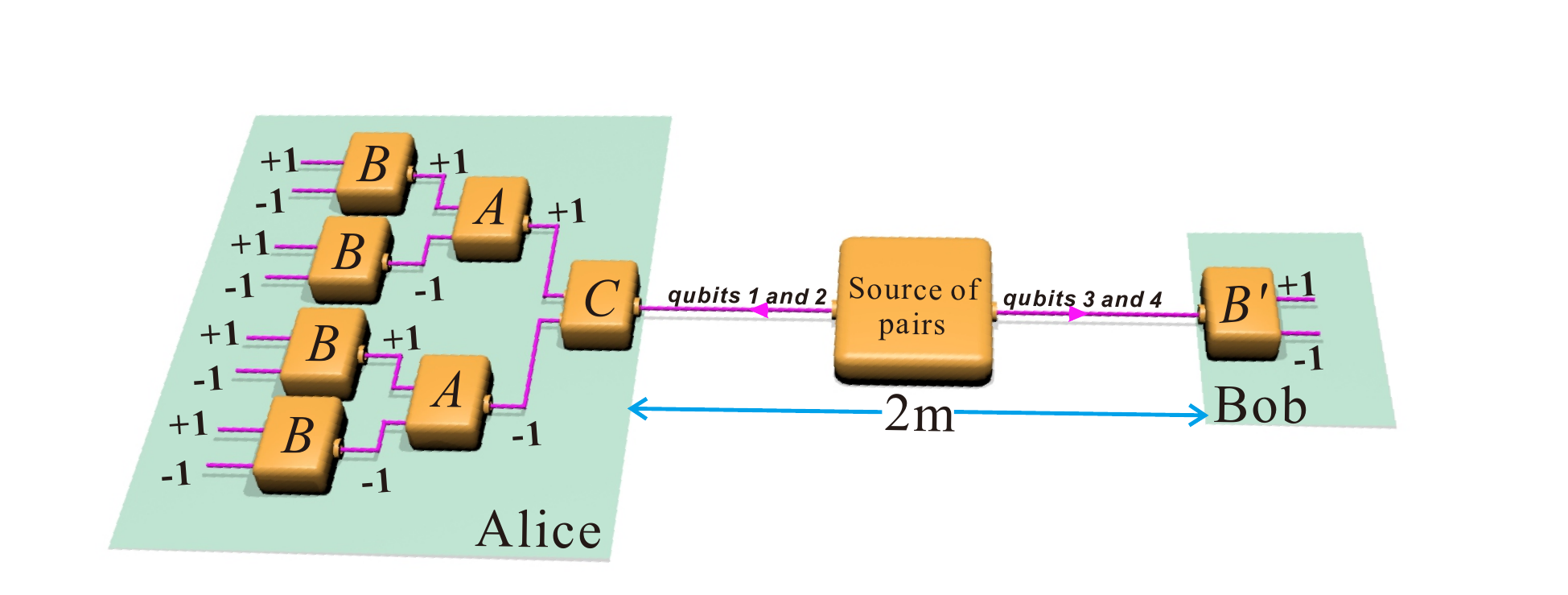}
\caption{
Scheme of the experiment. A source emits pairs of entangled particles. One particle encodes two qubits (qubits 1 and 2) and is sent to Alice's laboratory, where three sequential measurements are performed (e.g., $C$, $A$, and $B$). The other particle, encoding qubits 3 and 4, is sent to Bob's laboratory, where a single measurement is performed (e.g., $B'$). This allows us to obtain the correlations needed to test inequality (\ref{1}).
\label{Fig1}}
\end{figure}

%%%%%%%%%%%%%%%%%%%%%%%%%%%%%%%%%%%%%%%%%%%%%%%%%%%%%%%%%%%%%%%%%

{\em Experiment.---}In contrast to the standard ways of certifying nonlocality \cite{Bel64,FC72,ADR82,HBD15,GVW15,SMC15} and single-particle contextuality \cite{KZG09,ARB09,MRC10,ZUZ13,DHA13}, certifying nonlocality originated from single-particle contextuality requires observing the violation of an inequality for LHV theories (i.e., a Bell inequality which follows from the same assumptions under which any Bell inequality is valid), but made of correlations between sequential measurements on one particle and perfect correlations between some of these measurements and the corresponding measurements on a distant particle. The aim of our experiment is testing one of such inequalities proposed in Ref.\ \cite{Cab10} (see Supplemental Material for a derivation \cite{SM}),
\begin{equation}
\left\langle \omega \right\rangle \equiv \left\langle \chi \right\rangle
+\left\langle S\right\rangle \stackrel{\mbox{\tiny{LHV}}}{\leq} 16,
\label{1}
\end{equation}
where $\left\langle \chi \right\rangle$ only contains the correlations among the local successive compatible measurements on the first experimenter's (Alice's) particle (we will refer to them as Alice-Alice-Alice correlations) and $\left\langle S\right\rangle$ only contains the correlations between the measurements performed by Alice in the second or third place and the measurements performed by the second experimenter, Bob (we will refer to them as Alice-Bob perfect correlations). The fact that $\left\langle \omega \right\rangle$ only contains Alice-Alice-Alice and Alice-Bob perfect correlations is the distinctive signature of inequality (\ref{1}) with respect to standard Bell inequalities.

The correlations $\left\langle \chi \right\rangle $ and
$\left\langle S\right\rangle $ are defined as
\begin{align}
\left\langle \chi \right\rangle & =
\left\langle CAB\right\rangle +\left\langle cba\right\rangle +\left\langle \beta \gamma \alpha \right\rangle
+\left\langle \alpha Aa\right\rangle +\left\langle \beta bB\right\rangle -\left\langle c\gamma C\right\rangle,
\label{2} \\
\left\langle S\right\rangle & =
\left\vert \left\langle AA'\right\rangle_{CAB}\right\vert
+\left\vert \left\langle BB'\right\rangle_{CAB}\right\vert
+\left\vert \left\langle bb'\right\rangle_{cba}\right\vert
\notag \\
& +\left\vert \left\langle aa'\right\rangle _{cba}\right\vert
+\left\vert \left\langle \gamma \gamma '\right\rangle_{\beta \gamma \alpha}\right\vert
+\left\vert \left\langle \alpha \alpha '\right\rangle_{\beta \gamma \alpha}\right\vert
\notag \\
& +\left\vert \left\langle AA'\right\rangle_{\alpha Aa}\right\vert
+\left\vert \left\langle aa'\right\rangle_{\alpha Aa}\right\vert
+\left\vert \left\langle bb'\right\rangle_{\beta bB}\right\vert
\notag \\
&+\left\vert \left\langle BB'\right\rangle_{\beta bB}\right\vert
+\left\vert \left\langle \gamma \gamma'\right\rangle_{c\gamma C}\right\vert
+\left\vert \left\langle CC'\right\rangle_{c\gamma C}\right\vert, \label{3}
\end{align}
where $\left\langle CAB \right\rangle$ denotes the average of the product of the outcomes of $C$, $A$, and $B$ measured in that order, and $\left\langle BB'\right\rangle_{CAB}$ denotes the average $\left\langle BB'\right\rangle $ when Alice measures the ordered sequence $CAB$ and Bob measures $B'$. All measurements have two possible outcomes: $+1$ and $-1$.

Inequality (\ref{1}) can be derived from the assumptions of Conway and Kochen's free will theorem \cite{CK06,CK09}, but the inequality itself is independent of interpretational issues associated with the theorem, and we use it here to demonstrate that two-particle nonlocality can be produced from another, {\em a priori} different, simpler physical phenomenon, i.e., single-particle contextuality.

%%%%%%%%%%%%%%%%%%%%%%%%%%%%%%%%%%%%%%%%%%%%%%%%%%%%%%%%%%%%%%%%%
% Fig. 2
%%%%%%%%%%%%%%%%%%%%%%%%%%%%%%%%%%%%%%%%%%%%%%%%%%%%%%%%%%%%%%%%%

\begin{figure}[tb]
\centering
\includegraphics[trim = 1cm 0.5cm 0.5cm 3cm,clip,width=7.6cm]{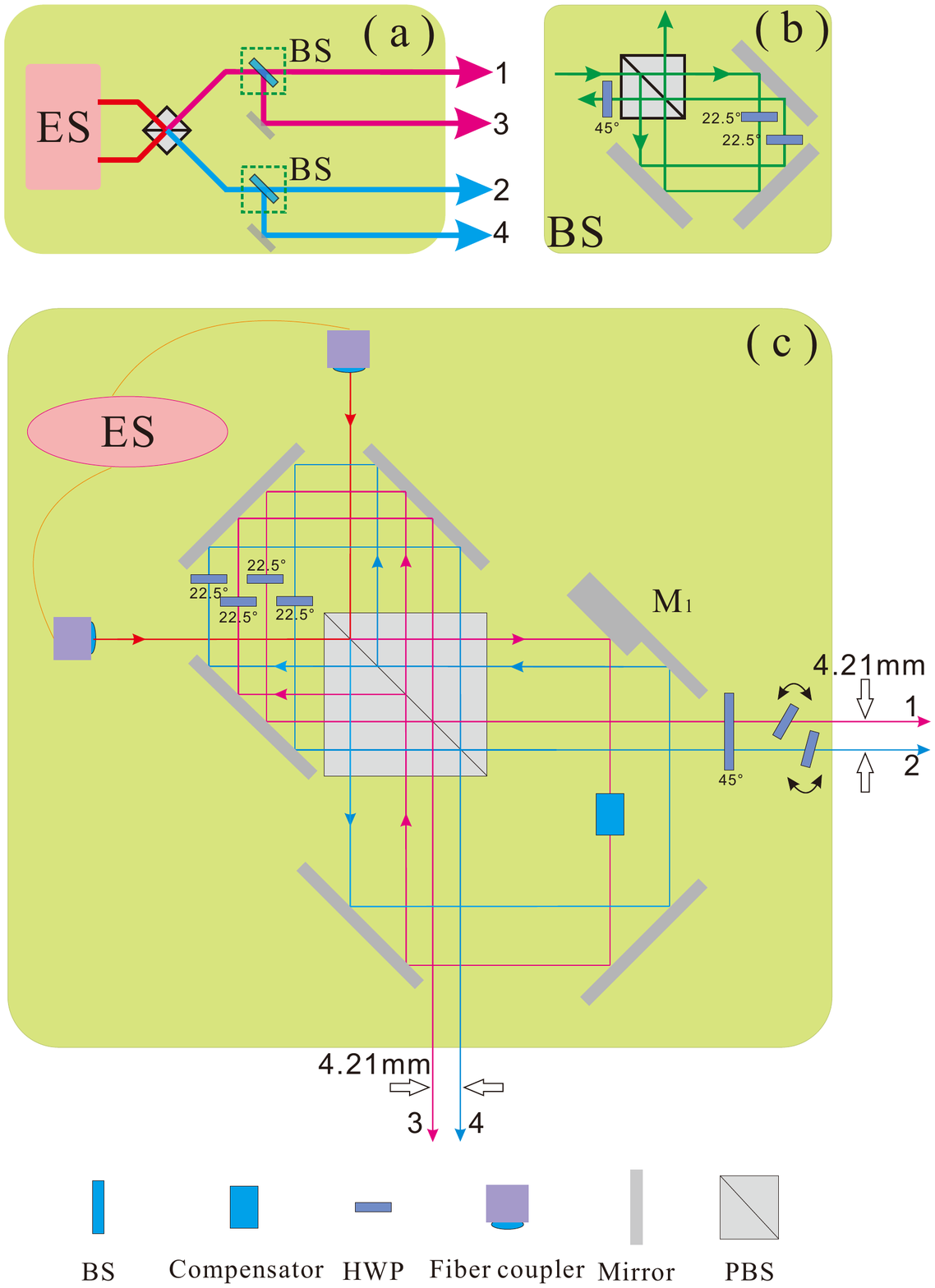}
\caption{
State preparation. (a) ES denotes the entanglement source and BS a specially designed beam splitter. (b) Specially designed beam splitter in detail. One photon is injected in a polarizing beam splitter (PBS), where it is reflected or transmitted depending on its polarization. Polarization is rotated by a half-wave plate (HWP) in both paths, which are then combined at the PBS. A HWP at $45^{\circ}$ is placed in one of the outputs of the PBS to rotate the state to the original one. If the two HWPs are set at $22.5^{\circ}$, then the whole device acts as a 50/50 beam splitter. (c) Experimental setup. ES prepares two photons in the state $\left\vert \protect\psi^{-}\right\rangle =\left( \left\vert H\right\rangle \otimes \left\vert V\right\rangle -\left\vert V\right\rangle \otimes \left\vert H\right\rangle \right) /\protect\sqrt{2}$ from two type-I cut $\protect\beta $-barium borate crystals. Both photons are injected into a PBS. A polarizing Hong-Ou-Mandel interferometer is adopted to remove the temporal and spatial distinguishability of the two photons. Then, the two photons are reflected by a special designed mirror $M_1$, which contains two agglutinate mirrors (the thickness of the left one is $2.968$ mm and the thickness of the right one is $6$ mm) to introduce a spatial separation of $4.21$ mm between two paths: the blue lines (corresponding to qubits 1 and 2) and the red lines (corresponding to qubits 3 and 4). The same beam displacer is also used in the measurement setup. An $8.232$-mm-thick glass is inserted into the red line as a phase compensator. The two photons are then separated by the specially designed BS. The resulting state is $\left\vert \Psi \right\rangle_{1234}=\left\vert \protect\psi^{-}\right\rangle_{13}\otimes \left\vert \protect\psi^{-}\right\rangle_{24}$.}
\label{Fig2}
\end{figure}

%%%%%%%%%%%%%%%%%%%%%%%%%%%%%%%%%%%%%%%%%%%%%%%%%%%%%%%%%%%%%%%%%

Our experiment is schematically illustrated in Fig.~\ref{Fig1}. Two hyperentangled photons (i.e., entangled in two different degrees of freedom) are distributed to two spatially separated laboratories, Alice's and Bob's. Alice receives qubits 1 and 2, encoded in, respectively, the spatial mode and the polarization of her photon, and performs three successive measurements on it. Bob receives qubits 3 and 4, encoded in, respectively, the spatial mode and the polarization of his photon, and performs a single measurement on it.

We prepare the two-photon four-qubit state $\left\vert \Psi\right\rangle_{1234} = \left\vert \psi^{-}\right\rangle _{13} \otimes \left\vert \psi^{-}\right\rangle _{24}$, where $\left\vert \psi ^{-}\right\rangle_{ij}=\left( \left\vert 0\right\rangle_{i} \otimes \left\vert 1\right\rangle_{j}-\left\vert 1\right\rangle _{i} \otimes \left\vert 0\right\rangle _{j}\right) / \sqrt{2}$ is the singlet state for qubits $i$ and $j$. For this purpose, we adopt the scheme shown in Fig.~\ref{Fig2}. A cw laser at $404$ nm pumps two $0.3$-mm-thick type-I cut $\beta $-barium borate crystals \cite{KMW95} to generate the two-photon two-qubit entangled state $\left\vert \psi^{-}\right\rangle=\left( \left\vert H\right\rangle \otimes \left\vert V\right\rangle -\left\vert V\right\rangle \otimes \left\vert H\right\rangle \right) /\sqrt{2}$, where $H$ and $V$ correspond to horizontal and vertical polarization, respectively. The experimental concurrence of this state was $0.995\pm 0.003$ (statistical errors only). Then, the two photons were sent to the Hong-Ou-Mandel interferometer with visibility $0.996\pm 0.001$ (statistical errors only) \cite{LHH15} and then directed to a polarizing beam splitter (PBS). The two photons leave the PBS together through the upper or lower ports and then were split by 50/50 beam splitters (BSs). Next, the spatial modes were postselected after a careful phase adjustment, so the final state is $\left\vert \Psi \right\rangle_{1234}$. The states $\left\vert 0\right\rangle$ and $\left\vert 1\right\rangle$ of qubit 1 (qubit 3) were encoded in the ``red'' paths 1 and 3 (``blue'' paths 2 and 4) in Fig.~\ref{Fig2}. The states $\left\vert 0\right\rangle $ and $\left\vert 1\right\rangle$ of qubits 2 and 4 were encoded in the $H$ and $V$ polarizations, respectively. We used a coincident count to discard all the events in which both photons are transmitted or reflected by the BSs. A phase stable Sagnac interferometer was adopted to construct the special BS \cite{GLY10} shown in Fig.~\ref{Fig2}(b). This BS has two advantages: it is polarization independent and its transmission/reflection ratio is controllable and can be set at nearly perfect $50/50$.

In our experiment, we had 2 m between Alice's and Bob's laboratories and we assumed that this prevents the information about Alice's (Bob's) measurement setting from arriving to the photons in Bob's (Alice's) laboratory. In addition, we tested that our experimental results are compatible with this assumption by checking that our results do not violate the no signaling principle (see Supplemental Material for details \cite{SM}).

The overall detection efficiency was 3.3\% and we assumed that the detected photons were a fair sample of the pairs emitted by the source. This assumption can be avoided by adopting high-efficient superconducting detectors and having excellent coupling from the source to the quantum channels and very low loss all the way from the source to the detectors (see Supplemental Material for a discussion of loopholes \cite{SM}).

%%%%%%%%%%%%%%%%%%%%%%%%%%%%%%%%%%%%%%%%%%%%%%%%%%%%%%%%%%%%%%%%%
% Fig. 3
%%%%%%%%%%%%%%%%%%%%%%%%%%%%%%%%%%%%%%%%%%%%%%%%%%%%%%%%%%%%%%%%%

\begin{figure}[tb]
\centering
\includegraphics[trim = 1.1cm 14cm 0cm 0cm,clip,width=8.2cm]{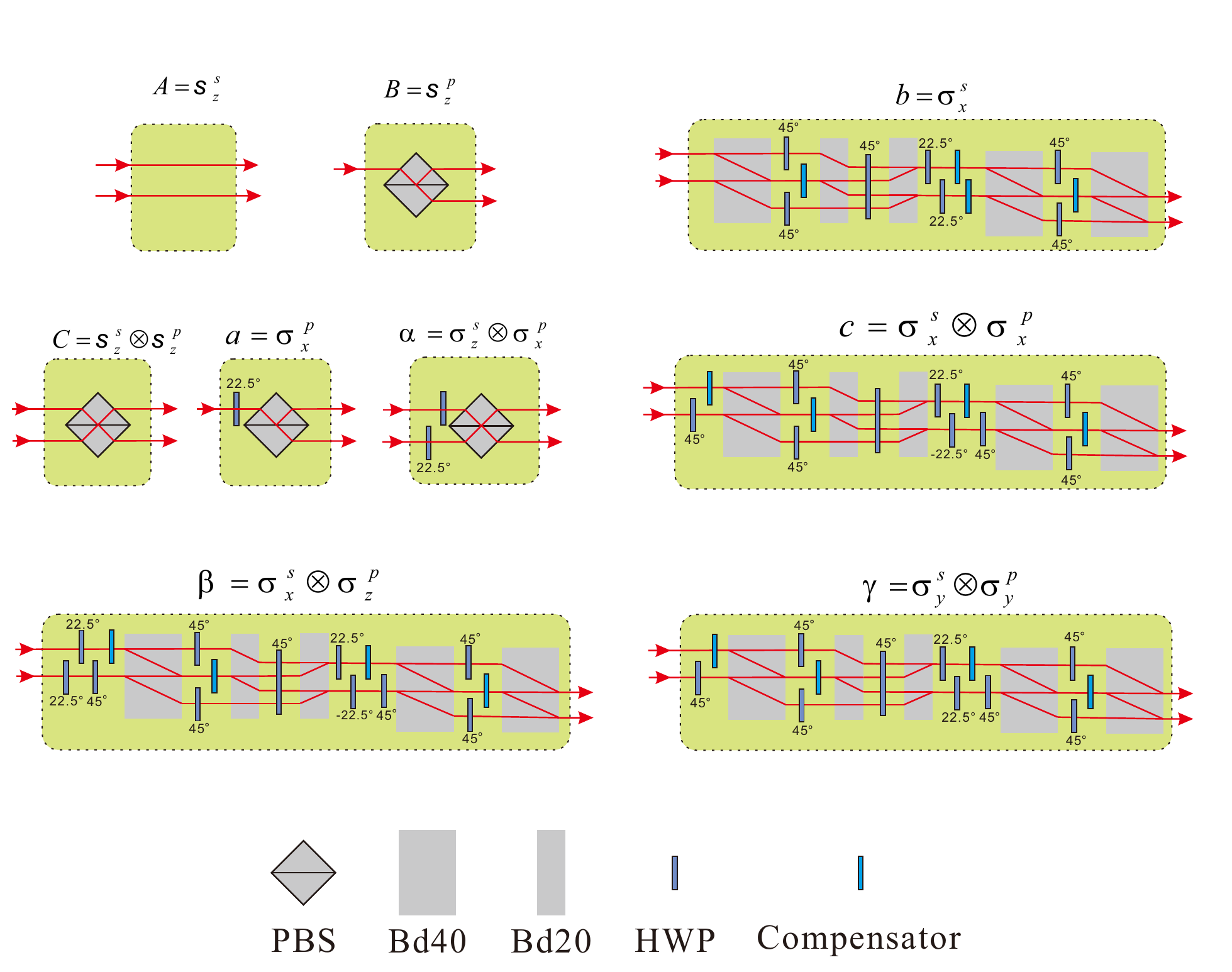}
\caption{
Devices for measuring the nine observables in Eq.\ (\ref{6}). Each device was made by combining polarizing beam splitters (PBSs), half-wave plates (HWPs), beam displacers (BDs) and thin glass plates that act as phase compensators. Here we describe the setup for measuring $b=\protect\sigma_{x}^{s}$. Consider that the input is its eigenstate $\left( \left\vert l\right\rangle+\left\vert r\right\rangle \right) /\protect\sqrt{2}$. This state is separated by the first Bd40 (a beam displacer that splits $H$- and $V$-polarized photons by $4.21$ mm at $808$ nm), depending on its polarization. To simplify the explanation, we will focus on the horizontally polarized part. Then, the $\left\vert l\right\rangle \otimes \left\vert H\right\rangle$ is reflected by the following Bd20 (a beam displacer that splits $H$- and $V$-polarized photons by $2.105$ mm at $808$ nm), while the $\left\vert r\right\rangle \otimes \left\vert H\right\rangle $ is rotated to $\left\vert r\right\rangle \otimes \left\vert V\right\rangle$ and transmitted by the first Bd20. Then, these two parts are rotated by a HWP and are combined together as $\left( \left\vert H\right\rangle +\left\vert V\right\rangle \right) /\protect\sqrt{2}$. Subsequently, it is rotated to $\left\vert H\right\rangle $ by a HWP and reflected by the second Bd40 to pass through another HWP, and then is rotated to $\left\vert V\right\rangle $ and transmitted by the last Bd40. The vertically polarized part is reflected by the last Bd40 and coherently aligned with the horizontally polarized part. This setup implements $b=\protect\sigma_{x}^{s}$ and does not affect the polarization degree of freedom.}
\label{Fig3}
\end{figure}

%%%%%%%%%%%%%%%%%%%%%%%%%%%%%%%%%%%%%%%%%%%%%%%%%%%%%%%%%%%%%%%%%

\begin{figure}[th]
\centering
\includegraphics[trim = 0.8cm 7.4cm 0 0.8cm,clip,width=8.2cm]{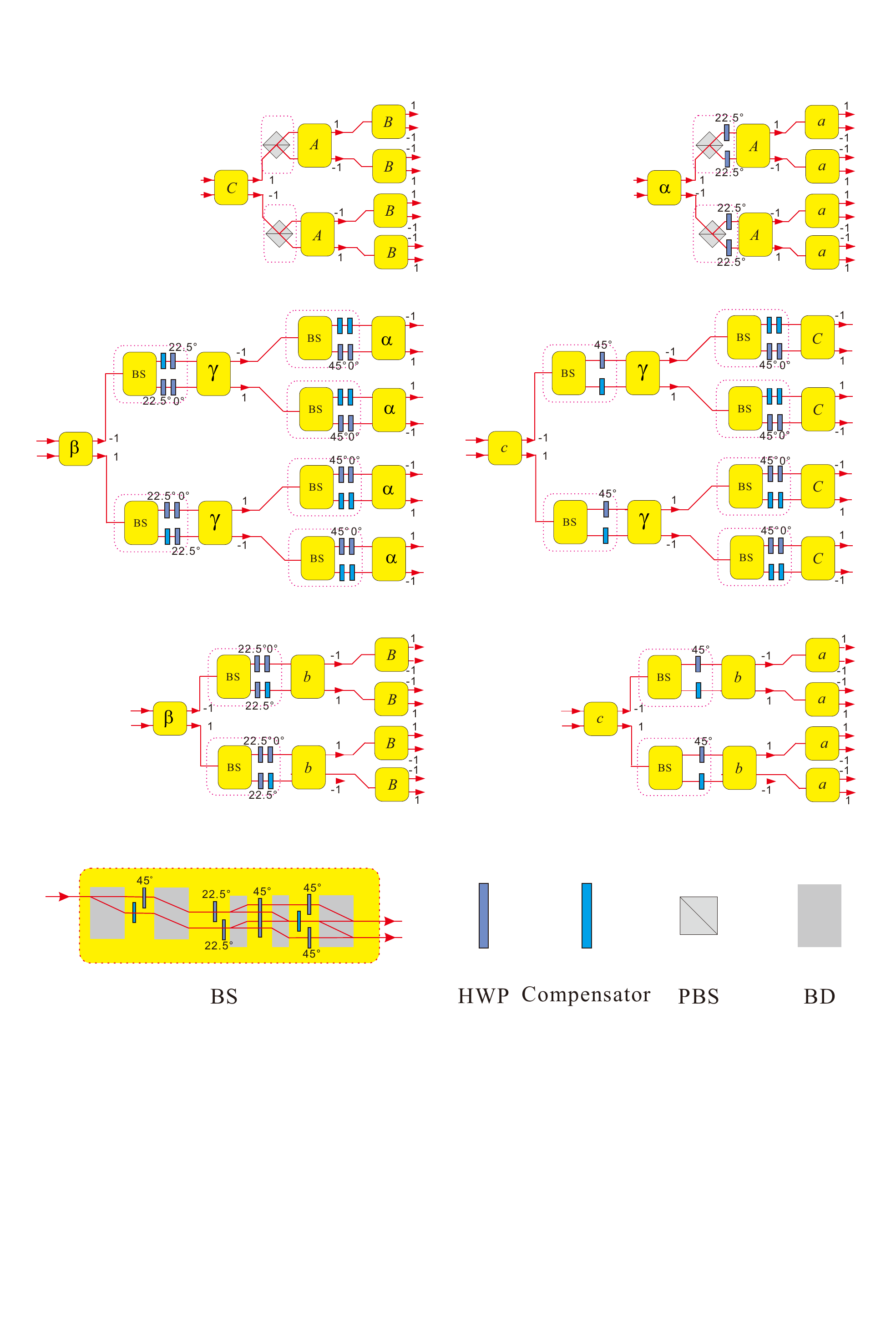}
\caption{
Setups for testing (\ref{2}). For measuring, e.g., $CAB$ we used the setup in the upper-left corner. There, one photon enters the device $C$ and exits at one of the two possible outputs depending of whether the outcome of $C$ is $+1$ or $-1$. Then the photon passes through a measuring device $A$ (for that, two identical measuring devices $A$ are placed at the outputs of $C$) and then through a measuring device $B$ (for that, four identical measuring devices $B$ are placed at the outputs of $B$) \cite{ARB09}. Hence, we can determine $\langle CAB \rangle$ by recording the photon counting probability after measuring device $B$ (from top to bottom named as $P_1,P_2,\ldots,P_8$, then $\left\langle CAB\right\rangle=P_1-P_2+P_3-P_4+P_5-P_6+P_7-P_8)$. The eigenstates of the measured observable are recreated before entering the next measurement, as our measuring devices map eigenstates to a fixed polarization and spatial mode. The specially designed BS [see Fig.~\ref{Fig2}(b)] was used to enable stable sequential measurements. The yellow boxes represent the setups shown in Fig.~\ref{Fig3}.}
\label{Fig4}
\end{figure}

{\em Test of contextuality.---}We tested contextuality, by testing the Peres-Mermin inequality \cite{Cab08}, which is valid for NCHV theories:
\begin{equation}
\left\langle \chi \right\rangle \stackrel{\mbox{\tiny{NCHV}}}{\leq} 4. \label{4}
\end{equation}
For this test, Alice measured six sequences: $CAB,$ $cba,$ $\beta \gamma \alpha ,$ $\alpha Aa,$ $\beta bB$, and $c\gamma C$, where
\begin{align}
A& =\sigma_{z}^{s},\text{ \ \ \ \ \ \ \ \ \ \ \ }B=\sigma_{z}^{p},\text{ \ \ \ \ \ \ \ \ \ \ }C=\sigma_{z}^{s}\otimes \sigma_{z}^{p}, \notag \\
a& =\sigma_{x}^{p},\text{ \ \ \ \ \ \ \ \ \ \ \ }b=\sigma_{x}^{s},\text{\ \ \ \ \ \ \ \ \ \ \ \ \ }c=\sigma_{x}^{s}\otimes \sigma_{x}^{p}, \notag \\
\alpha & =\sigma_{z}^{s}\otimes \sigma_{x}^{p},\text{ \ \ }\beta =\sigma_{x}^{s}\otimes \sigma_{z}^{p},\text{\ \ \ }\gamma =\sigma_{y}^{s}\otimes \sigma_{y}^{p}, \label{6}
\end{align}
and $\sigma_{x}^{i}$, $\sigma_{y}^{i}$, $\sigma_{z}^{i}$ denote the Pauli observables corresponding to the spatial mode ($i=s$) and polarization ($i=p$). Each of these nine observables \cite{Per90,Mer90} was measured using the devices shown in Fig.~\ref{Fig3}. The configurations corresponding to each of the six sequences are shown in Fig.~\ref{Fig4}. The designed beam displacer-based interferometer \cite{OPW04} had a visibility of approximately $0.998$ using an aligned laser source.

Our experimental result was
\begin{equation}
\left\langle \chi \right\rangle =5.817 \pm 0.011, \label{3}
\end{equation}
which violates the Peres-Mermin inequality (\ref{4}) by 165 standard deviations. To our knowledge, this is the largest value ever reported for the correlations of the Peres-Mermin inequality \cite{KZG09,ARB09,MRC10}. Detailed experimental results are provided in the Supplemental Material \cite{SM}.

%%%%%%%%%%%%%%%%%%%%%%%%%%%%%%%%%%%%%%%%%%%%%%%%%%%%%%%%%%%%%%%%%
% Fig. 4
%%%%%%%%%%%%%%%%%%%%%%%%%%%%%%%%%%%%%%%%%%%%%%%%%%%%%%%%%%%%%%%%%

%%%%%%%%%%%%%%%%%%%%%%%%%%%%%%%%%%%%%%%%%%%%%%%%%%%%%%%%%%%%%%%%%

{\em Test of perfect correlations.---}In the other laboratory, Bob chose among observables $A', B', C', a', b', \alpha'$, and $\gamma'$, which are identical to, respectively, $A, B, C, a, b,\alpha$, and $\gamma$ in Alice's laboratory (accent marks are just used to remind that these observables are measured on Bob's photons). For the state $\left\vert \Psi \right\rangle _{1234}$, observables $A$ and $A'$ are perfectly correlated so, by measuring one of them, an experimenter can predict with certainty the result of the corresponding measurement in the distant particle \cite{EPR35}. Similarly for $B$ and $B'$, $C$ and $C'$, $a$ and $a'$, $b$ and $b'$, $\alpha$ and $\alpha'$, and $\gamma$ and $\gamma'$. Consequently, the expected mean value is $\left\langle S\right\rangle =12$ for an ideal experiment. In our experiment we obtained
\begin{equation}
\left\langle S\right\rangle =11.430 \pm 0.016. \label{7}
\end{equation}
Detailed experimental results are provided in the Supplemental Material \cite{SM}. The difference with respect to the expected result is due to a nonperfect phase compensation in the state preparation. The value of $\left\langle S\right\rangle$ can be reproduced by LHV theories \cite{Bel66,KS67}. Therefore, Alice-Bob correlations, by themselves, do not reveal nonlocality.

%%%%%%%%%%%%%%%%%%%%%%%%%%%%%%%%%%%%%%%%%%%%%%%%%%%%%%%%%%%%%%%%%

{\em Test of nonlocality.---}However, when the experimental values of the Alice-Alice-Alice correlations are taken into consideration, then we observed that
\begin{equation}
\left\langle \omega \right\rangle =17.247 \pm 0.019, \label{8}
\end{equation}
which violates inequality (\ref{1}) by 66 standard deviations and therefore reveals Bell nonlocality.

%%%%%%%%%%%%%%%%%%%%%%%%%%%%%%%%%%%%%%%%%%%%%%%%%%%%%%%%%%%%%%%%%

{\em Conclusions.---}Our purpose has been to observe something which cannot be observed in any of the experiments testing simpler Bell inequalities \cite{Bel64,FC72,ADR82,HBD15,GVW15,SMC15}, namely, that two-particle Bell nonlocality can be produced from single-particle contextuality. From this perspective, the results of our experiment show that there are correlations in nature which cannot be explained by LHV theories {\em because} they contain single-particle correlations which cannot be reproduced with NCHV theories. This is revealed by the fact that the violation of inequality (\ref{1}), which proves nonlocality, can be traced back to the violation of inequality (\ref{4}), which proves single-particle contextuality. It is also revealed by the fact that the correlations between separated particles given by (\ref{7}), by themselves, admit an explanation in terms of LHV theories, while no such explanation is possible when single-particle correlations are taken into account. Therefore, from this perspective, our experiment shows a new way to produce nonlocality.

In addition, our experiment solves a problem that previous experiments testing the Peres-Mermin inequality \cite{KZG09,ARB09,MRC10} have. While the results of all these experiments can be simulated with classical models \cite{LaC09,Bla15,FBV16}, our experiment rules out all these models, since no contextual but local hidden variable model can explain the observed correlations. In this sense, our experiment constitutes a crucial development of the experiments in Refs.\ \cite{KZG09,ARB09,MRC10}.

From the perspective of quantum information, our experiment demonstrates that there is a connection between the two critical resources needed for, respectively, universal fault-tolerant quantum computation and device-independent secure communication. Moreover, our results show that both can be produced simultaneously by the same physical system. This is remarkable in light of recent results proving that this is impossible if we consider simpler forms of contextuality and nonlocality \cite{KCK14,JWG16,SR16,ZZL16}. Therefore, our experiment is also a proof of principle that quantum systems can be used simultaneously for quantum computation and secure communication.

Finally, our experiment can also be taken as a test of Conway and Kochen's free will theorem \cite{CK06,CK09}. Under the assumptions in Refs.\
\cite{CK06,CK09}, and modulo some loopholes, the violation of inequality (\ref{1}) implies that the results of the measurements on the photons are not determined by their past.

%%%%%%%%%%%%%%%%%%%%%%%%%%%%%%%%%%%%%%%%%%%%%%%%%%%%%%%%%%%%%%%%%%%

%\section*{Acknowledgments}

%%%%%%%%%%%%%%%%%%%%%%%%%%%%%%%%%%%%%%%%%%%%%%%%%%%%%%%%%%%%%%%%%%%

\begin{acknowledgments}
We thank A.\ J.\ L\'opez-Tarrida and A. Tavakoli for helpful comments. We are grateful to an anonymous referee for his or her efforts to help us improve the manuscript. This work was supported by the National Natural Science Foundation of China (Grants No.\ 11474267, No.\ 11274289, No.\ 11325419, No.\ 11374288, No.\ 11474268, No.\ 61327901, and No.\ 61225025), the Strategic Priority Research Program (B) of the Chinese Academy of Sciences (Grant No.\ XDB01030300), the Fundamental Research Funds for the Central Universities, China (Grants No.\ WK2470000018 and No.\ WK2470000022), the MINECO, Spain (Project No.\ FIS2014-60843-P, ``Advanced Quantum Information,'' with FEDER funds), the Knut and Alice Wallenberg Foundation, Sweden (Project ``Photonic Quantum Information''), and the FQXi (Large Grant ``The Observer Observed: A Bayesian Route to the Reconstruction of Quantum Theory'').
\end{acknowledgments}

%%%%%%%%%%%%%%%%%%%%%%%%%%%%%%%%%%%%%%%%%%%%%%%%%%%%%%%%%%%%%%%%%%%
%
% MINECO: Ministry of Economy and Competitiveness FundRef ID http://dx.doi.org/10.13039/501100001862
% FEDER: Fonds europ\'een de d\'eveloppement \'economique et r\'egional (European Regional Development Fund)
% Knut and Alice Wallenberg Foundation = Knut och Alice Wallenbergs Stiftelse FundRef ID http://dx.doi.org/10.13039/501100004063
%
%%%%%%%%%%%%%%%%%%%%%%%%%%%%%%%%%%%%%%%%%%%%%%%%%%%%%%%%%%%%%%%%%%%

%%%%%%%%%%%%%%%%%%%%%%%%%%%%%%%%%%%%%%%%%%%%%%%%%%%%%%%%%%%%%%%%%

\end{document}